\documentclass[sigconf]{acmart}

\acmYear{2026}

\usepackage{hyperref}
\usepackage{graphicx} 
\usepackage{enumitem} 
\usepackage{soul}
\usepackage{float}
\usepackage{listings}
\usepackage{framed}
\usepackage{xcolor}
\usepackage{color} 
\usepackage{framed} 
\usepackage[table,xcdraw]{xcolor}

\renewcommand\footnotetextcopyrightpermission[1]{} 
\makeatletter
\renewcommand\@copyrightpermission{}               
\makeatother

\makeatletter
\renewcommand\@copyrightpermission{}
\makeatother

\newenvironment{querybox}
  {\begin{framed}\begin{quote}\color{black}\small\itshape}  
  {\end{quote}\end{framed}}  

\date{}

\usepackage{todonotes}  

\begin{document}

\title{On Integrating Large Language Models and Scenario-Based Programming for Improving Software Reliability}

\author{Ayelet Berzack}
\email{ayelet.berzack@mail.huji.ac.il}
\orcid{0009-0009-9339-2202}
\affiliation{%
  \institution{The Hebrew University Of Jerusalem}
  \country{Israel}
}

\author{Guy Katz}
\email{g.katz@mail.huji.ac.il}
\orcid{}
\affiliation{%
  \institution{The Hebrew University Of Jerusalem}
  \country{Israel}
}

\begin{abstract}
Large Language Models (LLMs) are fast becoming indispensable tools for software developers, assisting or even partnering with them in crafting complex programs. The advantages are evident --- LLMs can significantly reduce development time, generate well-organized and comprehensible code, and occasionally suggest innovative ideas that developers might not conceive on their own. However, despite their strengths, LLMs will often introduce significant errors and present incorrect code with persuasive confidence, potentially misleading developers into accepting flawed solutions.

In order to bring LLMs into the software development cycle in a more reliable manner, we propose a methodology for combining them with ``traditional'' software engineering techniques in a structured way,
with the goal of streamlining the development process, reducing errors, and enabling users to verify crucial program properties with increased confidence. Specifically, we focus on the Scenario-Based Programming (SBP) paradigm --- an event-driven, scenario-based approach for software engineering --- 
to allow human developers to pour their expert knowledge into the LLM, as well as to inspect and verify its outputs. 

To evaluate our methodology, we conducted a significant case study, and used it to design and implement the Connect4 game. 
By combining LLMs and SBP we were able to create a highly-capable agent, which could defeat various strong existing agents. Further, in some cases, we were able to formally verify the correctness of our agent. Finally, our experience reveals interesting insights regarding the ease-of-use of our proposed approach. The full code of our case-study will be made publicly available with the final version of this paper.
\end{abstract}


\maketitle

\section{Introduction}
Since their appearance, large language models (LLMs) have dramatically changed the way complex software is designed and maintained. 
Modern LLMs, such as GPT-4o, are able to quickly produce high-quality code, and have become integral tools in the software engineering toolkit~\cite{DuLiWaWaLiChFeShPeLo24}. These models contribute to many aspects of the software life-cycle, including prototyping, debugging, deployment and  interpretability; and their use is likely to increase and expand in coming years~\cite{Ha25}.

Despite these impressive advantages, the rise in use of LLMs also presents unique challenges.
Most notably, LLMs often err --- for example, they might produce code that operates incorrectly in corner cases --- and they often do so with persuasive confidence. The increasing reliance of human engineers on these tools could thus lead to the deployment of faulty code, making the whole process potentially unreliable for complex, safety-critical  applications. Recent studies have already shown that LLMs sometimes perform poorly on certain tasks that involve interdependent logic, such as class-level code generation, where multiple methods must interact correctly~\cite{DuLiWaWaLiChFeShPeLo24}. Thus, novel methodologies for leveraging LLMs in a way that results in reliable software are sorely needed.

Our work here proposes one such new engineering methodology, aimed at mitigating the risks involved by using LLMs as part of the development cycle, while still retaining the benefits they afford.
Our proposed methodology is hybrid, in the sense that it combines human guidance with automated, LLM-based code generation. The idea is to allow human domain experts to pour their knowledge into the LLM in order to guide it, but at the same time ensure that the LLMs outputs are interpretable, and verifiable, by the domain expert.
This is achieved using structured prompts, iterative refinement, and human-in-the-loop feedback to reduce hallucinations and improve logical consistency.

One major challenge in this kind of approach is that it may be difficult for the human expert to assess the correctness of code generated by the LLM. To mitigate this difficulty, we advocate the use of Scenario-Based Programming (SBP)~\cite{HaMaWe12}, where complex software is expressed as a collection of inter-dependent scenarios of desirable or undesirable behavior. The advantages of using SBP in this context are two-fold. First, it allows for a well-structured way for the domain expert to incrementally refine the specifications provided to the LLM, until the desired outcome is reached. Second, scenario-based programs are known to be more readily interpretable, both by humans and by automated analysis techniques, such as model checking~\cite{MaHaHaMuTe18}; and this allows to more easily, and semi-automatically, inspect the LLM's outputs and gain confidence in their correctness.

Building on this foundation, we introduce a structured, SBP-based methodology for guiding LLMs in the development of complex software. The process begins with the developer defining high-level behavioral goals and decomposing them into discrete, modular scenarios, each representing a specific requirement or constraint. Prior to implementation, the LLM is provided with a preamble consisting of relevant background information --- both about the system being developed and about Scenario-Based Programming itself, which is not widely represented in existing online resources.
Implementation then proceeds incrementally: for each scenario, the developer engages in an iterative process with the LLM to generate a corresponding scenario object. This involves crafting focused, well-scoped queries that include relevant assumptions, context, and known events. The LLMs output is reviewed for correctness before integration --- typically through manual inspection, but sometimes supported by explicit or symbolic formal verification to ensure key properties are satisfied.
The LLM plays an active role throughout this process: it generates code, proposes refinements, and can even assist in debugging. When issues are discovered during testing or verification, the developer presents them to the LLM, which can help diagnose the problem and suggest or implement corrections. In this workflow, the LLM acts as a proactive coding assistant, while the developer guides the process, supplies domain knowledge, and ensures correctness.

\begin{figure}[h]
  \centering
  \includegraphics[width=0.5\linewidth]{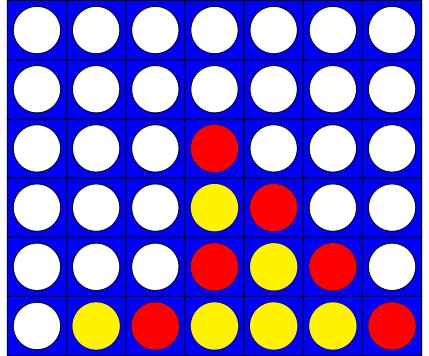}
  \caption{A Connect4 board game sample where red wins with a diagonal line.}
  \label{fig:connect4-sample}
\end{figure}

In order to evaluate our approach, we conducted an extensive case-study. Specifically, we created an agent for playing Connect4 --- a popular board game, where two players take turns dropping colored discs into a vertical $7\times 6$ board, aiming to be the first to form a line of four discs of their own color (typically red and yellow), either horizontally, vertically, or diagonally (Figure~\ref{fig:connect4-sample}). 
Connect4 is a partially-solved game: it has been formally proven that the first player can always win with perfect play~\cite{Al88}. 
However, existing perfect solutions rely on brute-force search, and it is presently unknown how to solve the $7 \times 6$ variant of the game using rules and heuristics comprehensible to humans. Thus, creating an agent whose behavior a human can readily understand remains an interesting problem, which we seek to tackle here.

The goal of our evaluation was thus to investigate whether strategic game-play could be effectively constructed by a large language model when guided by a structured SBP-based development process.
We sought to create, with the help of an LLM, a scenario-based agent that plays ``yellow'' in Connect4, and which achieves results comparable with state-of-the-art solutions. As part of our methodology, we first had the LLM generate scenarios that described the game rules; and later, also scenarios that implemented strategies. We repeatedly evaluated the agent's capabilities using both testing and formal verification, and iteratively query the LLM to modify and correct certain aspects of the scenarios. The modular nature of SBP facilitates this process, allowing the human-in-the-loop to focus on one particular aspect of the game each time. The LLM proved highly beneficial in translating natural language descriptions into working code --- specifically, turning scenario descriptions into scenario objects. This not only saved development time, but also led to the generation of novel code structures that we might not have considered independently. Our ability to explicitly verify the resulting program further strengthened the process: it enabled us to confirm that no rules were violated and provided formal guarantees that the agent would win under certain constrained move sequences.

Our agent achieved remarkable results, demonstrating the effectiveness of our hybrid methodology. It outperformed three distinct AI-based Connect4 opponents, all employing \emph{minimax} and \emph{alpha-beta pruning}, one of which boldly claims the title ``Unbeatable AI''.
The first is the ``Pro Player'' available at Tleemann’s Connect4 site~\cite{Tl25}, the second is the ``Unbeatable AI'' hosted at Websim~\cite{Do25}, and the third is an AI player developed by Keith Galli, available on his GitHub repository~\cite{Ga25}
 We even deployed our agent online\footnote{\url{https://connect4bpagent.pythonanywhere.com}} and invited users to try playing against it; members of our team found it difficult to defeat it.
In addition, although the agent is not always guaranteed to win, we used formal verification to prove that it will always win under several fixed sequences of the first \( x \) moves, providing valuable partial correctness guarantees. These results highlight the promise of combining LLM-driven code generation with scenario-based modeling to develop robust, maintainable, and high-performing systems.


The rest of this paper is organized as follows. In Section~\ref{sec:background} we provide necessary background on Large Language Models and Scenario-Based Programming. In Section~\ref{sec:methodology} we go into detail about our proposed methodology. In Section~\ref{sec:casestudy} we describe our case study and show how the methodology was applied to build a Connect4 agent. Section~\ref{sec:evaluation} presents a detailed evaluation of the agent’s performance, formal correctness, and development process. In Section~\ref{sec:relatedwork}, we discuss related work in the domains of LLM-based development and scenario-based modeling. We conclude and outline directions for future work in Section~\ref{sec:conclusions}.

\section{Background and Definitions}
\label{sec:background}
\subsection{Large Language Models (LLMs)}

\sloppy
Large Language Models (LLMs) are deep learning models trained on vast textual corpora to predict and generate coherent natural language. Recent LLMs, such as GPT-4~\cite{openai2023gpt4}, PaLM~\cite{chowdhery2022palm}, and Claude~\cite{An23}, exhibit impressive generalization capabilities. They can perform a wide range of tasks --- including natural language generation, translation, code synthesis, and debugging --- often with little or no task-specific training. This flexibility has made them powerful tools in software engineering workflows.

When applied to software development, LLMs can suggest code completions, generate functions from descriptions, refactor logic, and explain existing code. Their growing integration into IDEs and developer workflows has made them valuable assistants in everyday programming~\cite{cursorIDE}.

However, LLMs are inherently probabilistic and lack grounded semantic understanding. They may confidently generate incorrect or unexecutable code --- commonly referred to as \emph{hallucinations} --- and their outputs are often difficult to verify or formally constrain. These limitations raise concerns when using LLMs in safety-critical, logic-intensive, or formally verifiable domains, motivating hybrid approaches like the one explored in this paper.

\subsection{Scenario-Based Programming}
Scenario-Based Programming (SBP) is a programming paradigm in which the behavior of a system is specified in terms of interacting, independent scenarios, each representing a partial view of the system's functionality --- such as a requirement, a use case, or a constraint~\cite{HaMaWe12}. These scenarios are implemented as \emph{scenario objects} (also referred to as \emph{scenario threads}), which are executed concurrently and coordinate by declaring three types of event-related intentions: requesting events to occur, waiting for events to occur, and blocking events from occurring.

Execution in SBP progresses in discrete synchronization points. At each point, all active scenario objects announce their current declarations. The SBP runtime enviornment then selects one event to trigger --- among those that are requested by at least one scenario object, and not blocked by any. All scenario objects that requested or waited for the selected event then resume execution, while the others remain suspended; and the process then repeats at the next synchronization point.

\paragraph{Events.} An \emph{event} in SBP is an atomic action or occurrence in the system. Events are typically represented by simple data objects (e.g., a name or dictionary). Only events that are requested and not blocked can be selected for triggering by the engine.

\paragraph{Scenario-objects.} A \emph{scenario object} defines a self-contained behavior. Each scenario object yields control at synchronization points, specifying its interest in events. These objects are designed to be loosely coupled: they interact solely through their event declarations and are unaware of each other's internal state.

\paragraph{Synchronization points.} These are moments in the program's execution where all scenario objects pause and submit their event declarations. The runtime engine then selects a single event that satisfies the collective constraints. Once an event is selected, the relevant scenario objects proceed, maintaining event-driven coordination.

\paragraph{Behavioral composition.} The overall system behavior emerges as a composition of the independent contributions of all scenario objects. Because scenario objects are modular and communicate via events, developers can incrementally add or remove behaviors, often without needing to modify existing code~\cite{HaMaWe12}. This facilitates both maintainability and scalability.

In this paper, we use \emph{BPpy} --- a Python-based implementation of Scenario-Based Programming~\cite{Ya23} --- to illustrate the concepts and run simulations.
Below is a simple BPpy example that demonstrates the basic mechanics of SBP: two scenario objects requesting ``Hot'' and ``Cold'' water, respectively, and a third scenario object ensuring temperature stability, by forcing that hot and cold water be added alternately. The runtime selects and prints one event at a time:

\begin{lstlisting}
from bppy import *

@b_thread
def add_hot_water():
  while True:
    yield bp.sync(request=BEvent("Hot"))

@b_thread
def add_cold_water():
  while True:
    yield bp.sync(request=BEvent("Cold"))

@b_thread
def stabilize():
  while True:
    yield bp.sync(waitFor=BEvent("Hot"), 
                  block=BEvent("Cold))
    yield bp.sync(waitFor=BEvent("Cold"), 
                  block=BEvent("Hot))

bprog = SBProgram(
  bthreads=[add_hot_water(), add_cold_water(), 
            stabilize()]
)
bprog.run()
\end{lstlisting}

When this program runs, the two requested events are interleaved in an infinite loop. Because SBP relies on synchronization rather than imperative control flow, system behavior is the emergent result of collaborative event selection and thread progression.

\section{Methodology}
\label{sec:methodology}
In order to allow engineers to harness the capabilities of LLMs in a reliable manner, we propose the following methodology for designing complex software, using scenario-based programming. The methodology emphasizes incremental development, careful planning, and active collaboration between the developer and the LLM. The key insight is that while LLMs are effective in generating localized behaviors, they struggle with large-scale planning or modifying code holistically. Therefore, we recommend a human-guided, modular workflow, outlined below.

The workflow begins with an initial planning phase, where the developer defines the desired behaviors and ensures the LLM has the necessary background knowledge. Once this foundation is in place, the remaining scenario objects are implemented iteratively, one at a time, using a query–review–refine cycle until the system is complete.

\begin{itemize}
\item 
\textbf{Step 1: Define the Scenarios.}
Begin by thoroughly considering the overall goals and intended behavior of the program.
Go over the system specification, and for each requirement or rule, record it as an individual scenario. This process should feel intuitive, with no need to think about implementation details or code at this stage. Each scenario will later be implemented as a separate scenario object in the program.
Each scenario should have a narrow, well-defined responsibility. The more granular and modular the components, the easier they are to implement and verify.
Once a preliminary set is formed, present it (in free text) to the LLM, and ask whether it identifies any missing responsibilities. This process helps the LLM understand the scope of the system being developed and also helps the developer organize their thoughts. By the end of this step, there should be a clear, high-level structure for the SB program.

\item \textbf{Step 2: Provide Background Knowledge to the LLM.}
Before beginning the implementation, it is essential to equip the LLM with any relevant background knowledge that it may lack. This includes a general overview of Scenario-Based Programming, which is currently underrepresented in online resources --- resulting in limited familiarity on the part of the LLM compared to more mainstream programming paradigms.
It is also important to provide domain-specific background related to the application being developed, along with examples that can help guide the generation process.

\item\textbf{Step 3: Incremental Scenario-Thread Development and Refinement.}
 Begin implementation by asking the LLM to generate one scenario object at a time. For each scenario, provide a clear description of the required behavior and responsibility, including any assumptions or known events. Avoid asking the LLM to produce multiple components at once --- this often results in bloated, inconsistent, or confused output. After generating each scenario object, manually inspect the logic and correctness of the code before integrating it into the larger SB program. Because scenario objects are loosely coupled, this stepwise integration allows the system to be incrementally tested and debugged. If a scenario object does not meet the specification or requires adjustment, provide targeted feedback to the LLM and iterate on its response until the result is satisfactory.

\end{itemize}

\paragraph{Revising Previously Implemented Scenario Objects.}
Previously implemented scenario objects can be modified when necessary, either to refine behavior or to adapt to new requirements.
This process mirrors the approach used in Step~3: the developer queries the LLM with the existing scenario object and a description of the required changes.
When more extensive changes are needed --- such as modifying multiple scenario objects --- we found it significantly more effective to handle each change individually. Rather than asking the LLM to revise several scenario-threads at once, we achieved better results by querying it with a single code block and a focused instruction at a time. This approach aligns better with the model’s limitations and reduces the risk of introducing new errors.

\paragraph{Testing and Verifying.}
Throughout the process, treat the LLM as a coding partner rather than a fully autonomous agent. Frequently ask clarifying questions, present it with counterexamples, and challenge its assumptions. In return, the LLM will prove especially helpful in generating ideas, validating small design decisions, and drafting clean Python code. At any stage, if there is a property that should be verified, apply off-the-shelf verification tools for that individual object.

In summary, a successful collaboration with LLMs for BP development requires the developer to take an active planning role, decompose behaviors clearly, and work with the LLM incrementally. This results in modular, verifiable programs and avoids the common pitfalls of over-relying on the LLM to understand or construct complex architectures independently.

\section{Case Study: Connect4 Behavioral Program}
\label{sec:casestudy}
\subsection{Applying the Methodology: Implementing Core Game Mechanics}

In our case study we sought to address the following research questions:

\begin{itemize}
  \item \textbf{RQ1:} Can LLMs, when guided using structured prompts and SBP, generate correct and modular software components for complex systems?
  \item \textbf{RQ2:} Can an LLM-guided SBP agent achieve performance comparable to state-of-the-art agents in a non-trivial domain?
  \item \textbf{RQ3:} How convenient is it for an engineer to combine SBP and LLMs?
\end{itemize}

To that end, we applied our methodology to a complex case study: the task of generating a Connect4 game-playing agent. Our first goal was to create an agent that ``understands'' the game rules --- i.e., it plays valid moves, recognizes when the game ends, and adheres to the rules of the game. The second goal was to enhance this agent by implementing various strategies to improve its gameplay. In the remainder of this section, we demonstrate how our methodology helped us achieve the first goal, and in Section~\ref{subsec:strategy}, we focus on the second goal. We initially used OpenAI's GPT-4 (with a low-temperature setting) and later transitioned to Claude~\cite{An23} for longer sessions, taking advantage of its capabilities for more extensive development. Additionally, we incorporated the Cursor IDE~\cite{cursorIDE} for further flexibility and seamless integration during the longer sessions.

\paragraph{Step 1: Define the Scenarios.}

The SB program we seek to develop should represent a core Connect4 agent, which does not yet posses any game-playing strategies. Thus, the agent can select an arbitrary move in each turn, as long as it is allowed. We formulated the following scenarios:

\begin{itemize}
  \item \textbf{Board Representation.} The game board is a 7 by 6 grid, totaling 42 slots. The board is considered to be vertical (i.e., one side is the ``top''), and game discs from the top of a selected column and fall to the lowest available slot therein. Once a column is full, no more discs can be placed there.
  \item \textbf{Player Roles.} There are two players -- red and yellow --- each placing colored discs on their turn. Our agent plays yellow, and its opponent plays red. 
  \item \textbf{Turn Alternation.} Players alternate turns, starting with yellow.
  \item \textbf{Winning.} A player wins by placing four of their discs in a horizontal, vertical, or diagonal line.
  \item \textbf{Draw.} If the board is full with no winner, the game ends in a draw.
\end{itemize}

\paragraph{Step 2: Provide Background Knowledge to the LLM}
To initiate the implementation, we provided the chatbot with a preamble describing the Scenario-Based Programming paradigm, along with concrete examples. We instructed it to assist in building the Connect4 program using BPpy, while explicitly stating that it should not make assumptions about rules not provided. Since the game of Connect4 is well known, we did not need to provide detailed background; instead, we verified the LLMs familiarity and ensured that its understanding of the game's rules aligned with ours.

\paragraph{Step 3: Incremental Scenario-Thread Development and Refinement.}
In this key step, we began iteratively querying the LLM to produce the individual scenarios that represent the game rules. We instructed the LLM to implement a scenario object corresponding to each described behavior.
 We then reviewed the LLMs output. If the result was flawed, we iteratively refined the query and provided corrective feedback until it met our standards.

To illustrate this process, consider the implementation of the player-roles scenario for the red player. The initial query was:

\begin{querybox}
The second player is the user. On their turn, they input the column where they want to place a red disc. Just like the agent, they always try to place a disc in that column.
\end{querybox}

The LLM initially generated a scenario object \texttt{user\_player}, but it misunderstood the intended behavior.
Initially, the LLM attempted to determine the lowest available slot in the selected column and requested that specific placement. We corrected this by clarifying that this behavior was handled by other scenario objects. The \texttt{user\_player} object's role was simply to request all possible placements in the selected column. We also reminded the LLM to use a global list of red placement events. The corrected version is shown below:

\begin{lstlisting}
all_red_placements = [place_red(row, col) 
                      for col in range(num_cols) 
                      for row in range(num_rows)]

@bp.thread
def user_player():
  while True:
    col = int(input("Enter column number for 
                     placing red dics"))
    requests = [place_red(row, col) 
                for row in range(num_rows)]
    yield bp.sync(request=requests)
\end{lstlisting}


We repeated this process for each of the aforementioned game rules; and the LLM successfully  translated each of them into one or more scenario objects. For instance, the rule governing valid disc placement was assigned to a scenario object called \texttt{valid\_placement}, which is responsible for blocking illegal moves, such as placing a disc in a full column or at invalid row indices.

This example illustrates how the methodology fosters productive collaboration between the developer and the LLM: the developer defines precise responsibilities and verifies correctness, while the LLM generates the implementation within those constraints. For further details regarding the SB program design, please refer to Appendix~\ref{sec:appendixA}.

\subsection{Extending the System: Strategy Implementation and Verification}
\label{subsec:strategy}

Once the base game was functional, we moved on to the more challenging task of designing and implementing game strategies. We began by asking the LLM to list generally useful strategies, based on its background knowledge. We then prompted it to implement these strategies, one at a time. We started with relatively simple approaches, such as prioritizing the middle column and blocking immediate threats (opponent wins in the next move); and later  progressed to more complex strategies, like those involving fork intersections. As an additional source of background knowledge, we manually extracted useful strategies from the literature~\cite{Al25,Al88,Ga25}, and instructed the LLM to implement them.
As the strategies grew more complex, the LLM required more fine-grained guidance, but was still able to implement them successfully through well-scoped queries.

After implementing each batch of strategies, we periodically ran the verifier to identify scenarios where the red player could still win.
In one such instance, the yellow player missed an obvious win. We used this insight to query the LLM to implement win-detection scenario objects, which would prioritize placing a winning disc when three aligned yellow discs were already present.
Throughout this iterative process, we repeatedly:
\begin{enumerate}
  \item Ran the verifier to find counterexamples.
  \item Presented the red-win trace to the chatbot, and asked it how the current strategy could be improved.
  \item Confirmed the LLM's diagnosis (or manually diagnosed the problem if the LLM failed), and then queried the LLM to implement a solution.
\end{enumerate}

\begin{figure}[h]
  \centering
  \includegraphics[width=0.5\linewidth]{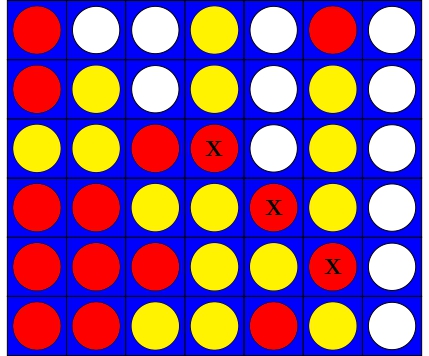}
  \caption{From the trace of a counterexample found by the verifier containing a red diagonal fork.}
  \label{fig:diagonal-fork}
\end{figure}

For example (Figure~\ref{fig:diagonal-fork}), in one case the verifier generated a trace with a diagonal fork: three red discs forming a diagonal with empty slots at both ends. Yellow could block only one, allowing red to win on the next turn. While the LLM could not identify the issue independently, we explained to it this new type of fork discovered and it readily implemented a fork-prevention strategy.

One useful observation is that the structure of SBP makes it straightforward to add test-cases as scenarios that force a particular trace (by blocking other events); and these scenarios can later be used to check whether the (undesirable) trace is no longer feasible. This allowed us to create a regression test-suite as we progressed.

\subsection{Extending SBP with new Features}\label{sec:features}

Throughout our work on the case study, we encountered two challenges that proved difficult to overcome using the existing SBP idioms and semantics. 

First, we observed that in some corner cases, certain strategies would block a potential move by the agent --- for example, because that move could lead to a later fork, and a consequent loss. However, the same move would actually complete a sequence of four discs, allowing the agent to win the game. Clearly, in such cases, we would prefer that the request to place the disc to be allowed, as the benefits outweigh the risks. However, according to SBP semantics, each scenario object is self-contained, meaning the object blocking the move is unaware that it should in fact be permitted. Additionally, under SBP semantics, blocking events takes precedence over requesting them. Once an event is blocked, there is no way for another scenario object to trigger it.

Secondly, during the development of the intersection fork scenario objects, we encountered situations where it was required to request events with different \emph{priorities} within a single synchronization point. For example, one case is a situation in which red is placing discs in a way that would simultaneously create two 3-long sequences of discs that could each be extended to winning, 4-long sequences. This is particularly risky when these two sequences intersect, and red places the intersection disc last --- so that yellow is able to block only one sequence in the next move, allowing red to win. When such a threat is detected, it can be prevented in advance by placing yellow discs that disrupt the two sequences of red discs at any of their locations; but the best solution is to capture the location of the linchpin, intersection disc before red places it. Thus, while several disc placements should be requested, the intersection disc should be requested with a higher priority.
Unfortunately, the standard SBP semantics have no notion of priorities; and the only existing extension with such a concept assigns priorities to entire scenario objects, and not to individual events~\cite{HaLaMaWe11}.

We resolved these issues by extending the SBP semantics to provide greater control over event handling. Specifically, we allowed each scenario object to assign an \emph{event-specific priority} to each event that it requests or blocks. The motivation to this idea (which was in fact proposed by the LLM when we presented the difficulty to it) was to allow blocked events to be triggered, provided that the request has a higher priority than the block. The event selection mechanism is then adjusted as follows.
An event is selected for triggering if:
\begin{itemize}
    \item It is requested by at least one scenario object.
    \item If it is blocked, it is requested with a higher priority than the priority with which it is blocked. 
    \item It is requested with the highest  priority among all events that meet the above criteria.
\end{itemize}

In order to be compatible with SBP's original blocking mechanism and the advantages it affords, we allowed blocked events to be blocked with a priority of $\infty$; alternatively, this could be thought of as simultaneously allowing ``hard blocking'' of events that can never be triggered, and ``soft blocking'' of events that may be triggered if they are requested with a higher priority than that with which they are blocked. Clearly, an event that is hard-blocked can never be triggered.

For further details on the precise semantics of these extensions to SBP, which may have broader applications, we refer the reader to Appendix~\ref{sec:appendixB}.

\section{Evaluation}
\label{sec:evaluation}

To evaluate our hybrid Connect4 agent, 
we measured its robustness and capabilities using formal analysis (addressing \textbf{RQ1}), and also pitted it against external AI-based Connect4 agents used as benchmarks (addressing \textbf{RQ2}). Finally, 
we documented and assessed our experience interacting with the LLM throughout development (addressing \textbf{RQ3}). Naturally, the first two aspects are easier to quantify, whereas the third is more subjective. 

\subsection{Agent Performance}

\begin{figure}[h]
  \centering
  \includegraphics[width=0.5\linewidth]{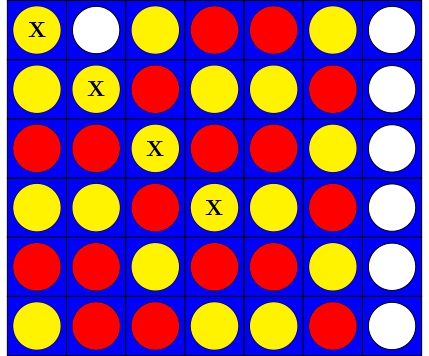}
  \caption{SBM Agent (yellow) vs AI Pro (red).}
  \label{fig:winning-state-1}
\end{figure}

\begin{figure}[h]
  \centering
  \includegraphics[width=0.5\linewidth]{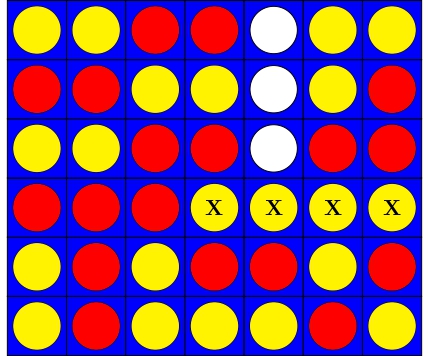}
  \caption{SBM Agent (yellow) vs ``Unbeatable AI'' (red).}
  \label{fig:winning-state-2}
\end{figure}

\begin{figure}[h]
  \centering
  \includegraphics[width=0.5\linewidth]{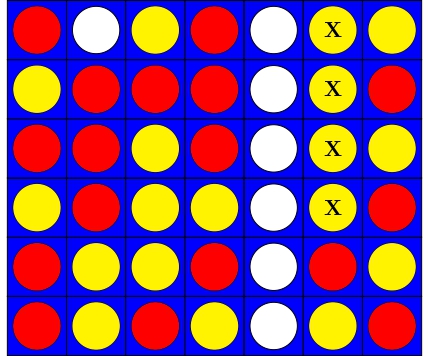}
  \caption{SBM Agent (yellow) vs ``Galli's AI'' (red).}
  \label{fig:winning-state-3}
\end{figure}

We  evaluated our agent against three strong, AI-based opponents available online: the ``Pro Player''~\cite{Tl25}, ``Unbeatable AI''~\cite{Do25}, and the AI player by Keith Galli~\cite{Ga25}. In each case, our agent played as the first player (yellow). We note that our agent is deterministic, and so are ``Unbeatable AI'' and the Keith Galli agent; and so repeating the experiment simply repeated the same game. Conversely, the ``Pro Player'' has some small degree of non-determinism, and so we had our agent play 10 matches against it. \emph{Our agent was able to win every single match, against all opponents, with no losses or draws.}
We regard this as indication that our agent implements a reliable and successful strategy. Figures~\ref{fig:winning-state-1}, \ref{fig:winning-state-2}, and \ref{fig:winning-state-3} show the final board states from matches against the three opponents.

We thus conclude that the answer to \textbf{RQ2} is positive: our results demonstrate that an LLM-guided SBP agent can repeatedly defeat advanced opponents.

\subsection{Formal Guarantees}

To complement gameplay testing, we used BPPy's explicit model checker (\texttt{DFSBProgramVerifier}) to verify partial correctness. The current state-of-the-art in BP model checking is insufficient for covering the large state space induced by Connect4; and so, in order to be able to at least partially verify our agent, we studied six \emph{fixed} opening sequences (i.e., a set of fixed initial moves by both players). Restricting the game to start from such a sequence shrunk the search space, and allowed the BPPy model checker to scale to the problem. From each such configuration, we successfully verified that our agent (yellow) was guaranteed to win, regardless of the subsequent moves played by its opponent. 

\begin{figure}[h]
  \centering
  \includegraphics[width=0.5\linewidth]{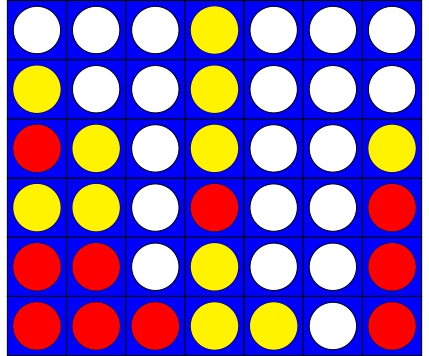}
  \caption{Fixed Moves Example 1.}
  \label{fig:fixed-seq-1}
\end{figure}

Figure~\ref{fig:fixed-seq-1} illustrates one such opening sequence, where we observe that yellow has a diagonal odd-threat (the last disc that needs to be placed is in an odd row, and it will inevitably be yellow's turn when it is time to place it); and also that yellow controls most of the middle column. Both of these are strategies employed by our agent, and so such a sequence of moves is quite plausible in a real match.

\begin{figure}[h]
  \centering
  \includegraphics[width=0.5\linewidth]{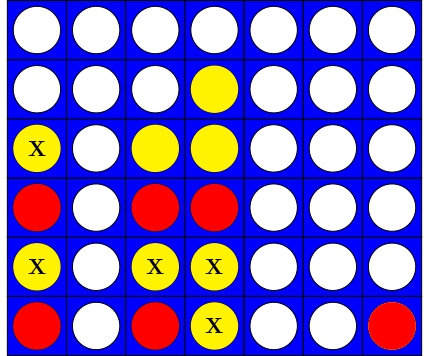}
  \caption{Fixed Moves Example 2.}
  \label{fig:fixed-seq-2}
\end{figure}

Figure~\ref{fig:fixed-seq-2} shows another verified opening configuration, in which yellow forms a classic \emph{fork}, simultaneously creating a diagonal and a horizontal threat. The discs marked with an “X” highlight positions where yellow is poised to win via either line, making it impossible for red to defend against both. This configuration, which occurs in practice due to our agent's strategies, 
demonstrates the agent’s ability to construct complex, multi-threat configurations through modular, verifiable logic.

\begin{figure}[h]
  \centering
  \includegraphics[width=0.5\linewidth]{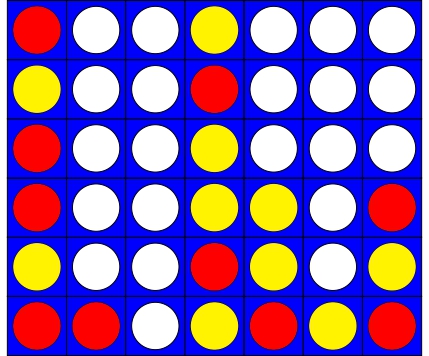}
  \caption{Fixed Moves Example 3.}
  \label{fig:fixed-seq-3}
\end{figure}

Finally, Figure~\ref{fig:fixed-seq-3} presents yet another verified game state, in which yellow is guaranteed to win --- as confirmed by the model checker. While the victory is not as immediately apparent from the board state compared to the previous examples, it is clear that yellow’s strategic control over the center column plays a critical role in securing the win. This configuration, again reachable by our agent, highlights the advantage of using formal verification to prove that the agent is guaranteed to win from a certain board state, even when the path to victory is not as visually obvious.

These formal results indicate a positive answer to \textbf{RQ1}, showing that the LLM-generated code is verifably correct in many cases. While the agent is not perfect --- it may lose, as some edge cases remain unhandled --- we demonstrate that a significant subset of the strategy is provably correct under formal analysis.

\subsection{Developer Experience}

Developing the Connect4 game was both an exciting and challenging experience. Initially, our expectations for the LLM were high --- we anticipated it would generate near-perfect scenario-based code. However, we quickly realized this was not the case, partially due to the limited information and examples available on SBP, and especially concerning the BPPy library.

In the early stages of the case study, we became slightly frustrated with the LLM's tendency to hallucinate functionality. These errors were challenging to overcome, but we soon recognized that providing clear feedback allowed the model to improve rapidly. For instance, the draw-detection logic was generated almost correctly on the first try once we supplied the appropriate context. With iterative refinement and formal verification, we were able to identify and fix errors effectively. One example involved a counterexample (a game trace in which the red player defeats our agent), which revealed a vulnerability related to a diagonal fork. After presenting this issue to the LLM, it successfully generated a mitigation scenario.

A particularly valuable aspect of the LLM was its ability to offer creative solutions when we felt stuck. It was exciting to witness how seamlessly the game strategies could be translated into scenario objects in our SB program. Progress was evident not only as we played against the agent, but also as we ran simulations, compared the agent’s performance with online agents, and validated the system using formal verification. Though there were moments where it felt like we were wasting time with seemingly unproductive conversations with the LLM, we found that when we adhered to the methodology --- focusing on one scenario object at a time with precise and clear prompts --- the results were highly effective.

In retrospect, we feel confident that future development will be more efficient, thanks to the refined methodology and deeper understanding that we have gained from this experience. Overall, the LLMs were highly effective at transforming behavioral descriptions into code, significantly accelerating the development process. Furthermore, SBP’s modularity played a crucial role in isolating logic and supporting verification, which contributed to a more structured and manageable workflow. Thus, with respect to \textbf{RQ3} we believe that there is a certain learning curve that must be mastered in order to make the most of our proposed methodology; but that once that phase is reached, it becomes quite convenient and enjoyable. Naturally, much additional experimentation is still needed to fully establish this claim in a broader manner.

\section{Related Work}
\label{sec:relatedwork}

Recent developments in Large Language Models (LLMs) have inspired a surge of interest in using generative models to assist with software engineering tasks. Several works have studied the effectiveness of LLMs in generating program code from natural language prompts~\cite{DuLiWaWaLiChFeShPeLo24,Ha25}. However, research increasingly emphasizes that unstructured use of LLMs often results in unreliable or incorrect code. To address this, recent work has proposed frameworks that integrate LLMs into structured development pipelines. For example, Sun et al.~\cite{SuShPaBa24} introduce \textit{Clover}, which adds verification to LLM-generated code using a closed-loop feedback cycle. Similarly, Yixuan et al.~\cite{LiPaPo24} present a framework in which LLMs guide enumerative program synthesis through specification refinement and ranking of candidate implementations.
Both emphasize the need for structured workflows when deploying LLMs in safety --- or correctness-critical domains.

Our work aligns with recent efforts to structure LLM-based development using scenario-based techniques~\cite{HaKaMaSz24}. The motivation for focusing SBP is that scenario objects tend to be well-aligned with how humans perceive systems, as well as amenable to formal analysis techniques~\cite{HaKaMaWe15}.

The connection between LLMs and scenario-based programming has previously been explored by Yaacov et al.~\cite{YaElWe24}, who argued that structuring code generation around modular scenario threads improves verifiability and traceability. They show how each requirement can be transformed into an independent code module, and how SBP modularity can guide prompt engineering and reduce LLM-induced errors, using both explicit and symbolic verification. Building on this work, we take their research a step further by creating a methodology for combining LLM development with SBP. We demonstrate how this methodology can be applied to develop more complex systems, providing a comparative analysis with existing systems available online. Additionally, we demonstrate how performing formal verification on the complex systems under development can help ensure correctness in this setting. Finally, we introduce new features to the SBP semantics (and the BPPy package), facilitating its integration with LLMs and its use in devising complex systems. 

On a broader scope, the usefulness and applicability of scenario-based
programming has been demonstrated in various contexts, such as program
repair~\cite{HaKaMaWe12,HaKaMaWe14,Ka21}; compositional and symbolic
verification and analysis~\cite{HaKaKaMaMiWe13,
  Ka13,HaKaLaMaWe15,KaBaHa15,GeGrGuKoGlMaKa17,KaMaSaWe19,KaEl21,HaKa14};
automated
optimization~\cite{HaKaKa13,HaKaKaMaWeWi15,HaKaMaMa16,GrGrKaMaGlGuKo16,HaKaMaMa16b,StGrGrHaKaMa17,HaKaMaMa18,StGrGrHaKaMa18,HaKaMaSaWe20};
synthesis~\cite{GrGrKaMa16}; and in the context of deep
learning~\cite{Ka20,Ka21b,YeAmElHaKaMa22,YeAmElHaKaMa23,AsKa23,CoYeAmFaHaKa24,HaKaMaSz25,YaWeAsKaZi25,AhStElKa25}.
  
Our agent is a scenario-based system constructed from declarative rules and domain-specific strategies. Some of these strategies were inspired by classical work in game-solving, most notably Allis’ seminal thesis~\cite{Al88}, which proves that Connect4 is a solved game where the first player can force a win using a combination of brute-force search and domain-specific heuristics. However, to our knowledge, no prior work has attempted to build a non-losing Connect4 player using only modular scenario-based rules combined with LLM-assisted development, nor to verify such an approach using formal methods.

\section{Conclusion and Future Work}
\label{sec:conclusions}

Large Language Models (LLMs) have recently gained widespread popularity among developers for their ability to generate or assist in generating code. However, despite their utility, LLMs are often unreliable, difficult to verify, and struggle to produce correct behavior in complex systems. In this paper, we proposed a novel methodology that addresses these challenges by integrating LLMs with Scenario-Based Programming (SBP). Our approach enables developers to leverage the strengths of LLMs --- such as code generation and strategic reasoning --- while mitigating their weaknesses through structured, modular, and verifiable design.
By using SBP as a foundation, we facilitate an incremental development process in which LLMs contribute to individual scenario components rather than attempting to construct an entire system at once. This is enabled by SBP’s modular architecture, where scenario objects operate independently. Such structure not only aligns well with the strengths of LLMs but also supports rigorous formal verification. It enhances the robustness and comprehensibility of the resulting code, making it easier to identify and correct errors. Ultimately, our approach demonstrates how LLM-assisted development can be carried out in a more controlled, reliable, and scalable manner. We demonstrated the effectiveness of the approach using the Connect4 game as a case-study.

Future research may focus on several key directions. One avenue is the development of more scalable and efficient model checking techniques tailored to SBP, enabling faster analysis of larger or more dynamic systems. Another is the application of our methodology to additional domains beyond games, such as robotics, smart environments, and human-computer interaction. Finally, we envision the creation of LLM agents specifically adapted for scenario-based reasoning—potentially by fine --- tuning on SBP artifacts or incorporating behavioral constraints during generation --- to further improve coherence, safety, and reusability within scenario-driven development.

\appendix

\section{Appendix: Program Structure}\label{sec:appendixA}
The Connect4 SB program is implemented as a modular system, organized into several key components, each responsible for a specific aspect of the game logic. Following scenario-based programming principles, these components are implemented as independent scenario objects that synchronize on shared events to drive the game forward.

\subsection{Game Events}

In our implementaiton, events are defined in a \texttt{GameEvents} class and precomputed at load time for efficiency. They fall into two main categories:

\paragraph{Placement Events.} These events represent player moves and carry positional data:
\begin{itemize}
    \item \texttt{PlaceYellow(\{'row': r, 'col': c\})}
    \item \texttt{PlaceRed(\{'row': r, 'col': c\})}
\end{itemize}

\paragraph{Special Events.} Used to control game flow:
\begin{itemize}
    \item \texttt{RequestUserInput}: Signals a prompt to the user.
    \item \texttt{Win(\{'color': 'Y' or 'R'\})}: Signals game termination on win.
    \item \texttt{Drae}: Signals game termination on draw.
\end{itemize}

For convenience, the \texttt{GameEvents} class also provides:
\begin{itemize}
    \item Lists of all red/yellow placement events.
    \item Aggregated event lists (e.g., all non-win events).
\end{itemize}

\subsection{Scenario Objects}

The core game mechanics are implemented using several foundational scenario objects. In the remainder of this Appendix, we provide function signatures for these objects to give the reader an overview of their structure. The full code  will be made publicly available with the final version of this paper.

\subsubsection{Basic Game Management}

The following scenario objects manage different aspects of the core game, such as checking for win conditions and handling player turns.

First, board management:

\begin{lstlisting}
@bp.thread
def board_manager():
  """Manages the game board state and visualization"""
  ...
\end{lstlisting}

Next, the enforcement of alternating turns between players:

\begin{lstlisting}
@bp.thread
def enforce_turns():
  """Enforces alternating turns between Yellow and Red in interactive mode"""
  ...
\end{lstlisting}

The following scenario checks for horizontal winning conditions. Similar scenario objects are used to check for vertical and diagonal winning sequences:

\begin{lstlisting}
@bp.thread
def check_horizontal_win(row, start_col, color):
  """Detects horizontal winning sequences."""
  ...
\end{lstlisting}

For detecting a draw situation:

\begin{lstlisting}
@bp.thread
def check_draw():
  """Monitors the game for a draw condition (full board)"""
  ...
\end{lstlisting}

The handling of winning conditions:

\begin{lstlisting}
@bp.thread
def handle_win():
  """Handles win events by stopping the game"""
  ...
\end{lstlisting}

The agent's behavior:

\begin{lstlisting}
@bp.thread
def computer_player():
  """Yellow (computer) player that can place pieces in any valid position"""
  ...
\end{lstlisting}

The user player's input handling:

\begin{lstlisting}
@bp.thread
def user_player():
  """Human player interface for Red pieces"""
  ...
\end{lstlisting}

\subsubsection{Strategies}

The agent’s strategic layer is guided by priorities, stored in a centralized class, which helps determine which strategies are executed based on the game state. The following signatures illustrate some of the strategies used:

\begin{lstlisting}
class Priorities:
...
BLOCK_BELOW_THREAT = 19800
BLOCK_THREAT = 19900
WIN = 20000
\end{lstlisting}
When assigning priorities for blocking or completing a threat, we also take into account other factors, such as parity --- i.e., whether a slot is in an even or odd row. Parity must be considered with respect to the alternating turn sequence, to determine whose turn it will be when a slot becomes available.

\paragraph{Win Threats}

The following strategy detects winning horizontal threats:

\begin{lstlisting}
@bp.thread
def win_horizontal_threat(horizontal_line):
"""
Requests horizontal winning moves for Yellow.
Starts requesting with lower priority when 2 pieces are detected and slots below are filled
(or immediately for bottom row), then increases priority when 3 pieces are present.
Implements the "playing the odd rows" strategy by prioritizing potential wins in odd-numbered rows
(rows 1, 3, and 5) with a higher priority (POTENTIAL_WIN_P1).
Args:
row (int): The row of the potential win
start_col (int): The starting column of the potential win
"""
...
\end{lstlisting}
The same approach is applied to vertical and diagonal winning threats, with corresponding functions for each.

\paragraph{Block Threats}
The following strategy detects and block potential horizontal red threats.

\begin{lstlisting}
@bp.thread
def block_horizontal_threat(horizontal_line):
"""
Blocks potential horizontal winning threats by Red.
Starts blocking when 2 pieces are detected and slots below are filled,
with increased priority when 3 pieces are present.
Implements a counter-strategy against Red's "play the even rows" strategy
by prioritizing blocking threats in even-numbered rows with a higher priority.
Args:
row (int): The row of the potential threat.
start_col (int): The starting column of the potential threat.
"""
...
\end{lstlisting}
The same approach is applied to vertical and diagonal blocking threats, with corresponding functions for each.

\paragraph{Fork Management}
Forks are critical situations where an opponent can create two simultaneous winning threats, making it impossible to block both. We identified several types of forks --- some known before development, and some discovered during verification when red beat our agent. To address these, we implemented scenarios that manage both red and yellow forks. These scenarios request that yellow place discs to block red's forks, and that yellow create its own forks.
For example, here is a scenario object that manages red horizontal forks:
\begin{lstlisting}
@bp.thread
def manage_red_horizontal_fork(line_of_five):
"""
Monitors a horizontal line of 5 slots for potential Red fork situations and blocks them.
Blocking priority increases based on how many slots below are filled.
Implements the parity strategy by considering the row number of the horizontal line
to determine priority. For Red's forks, we want to prioritize blocking forks in even rows.
Args:
line_of_five: List of 5 positions [(row, col), ...] forming a horizontal line
"""
...
\end{lstlisting}


\paragraph{Additional Strategies}
Finally, an important strategy is the prioritization of disc placement in the center column.

\begin{lstlisting}
@bp.thread
def center_column_strategy():
"""Prioritizes placing pieces in the center column"""
...
\end{lstlisting}

\section{Appendix: Extensions to BPpy}\label{sec:appendixB}

As described in Section~\ref{sec:features}, we introduced an extension that enables scenario objects to assign a \texttt{requestPriority} to each event it requests and a \texttt{blockPriority} to each event it blocks. The mechanism works as follows:

A single sync statement can include the following list parameters:
\begin{itemize}
  \item \textbf{waitFor}: A list of events the scenario object is waiting for. If one of these events is selected, the scenario object proceeds in the code until the next yield statement.
  \item \textbf{request}: A list of events that will be requested with \texttt{requestPriority}. Similar to \texttt{waitFor}, if one of these events is selected, the scenario object continues to the next step.
  \item \textbf{softBlock}: A list of events that will be blocked with \texttt{blockPriority}.
  \item \textbf{hardBlock}: A list of events that will be blocked unconditionally, regardless of any priority.
  \item \textbf{requestPriority}: A single priority value assigned to all events in the \texttt{request} list.
  \item \textbf{blockPriority}: A single priority value assigned to all events in the \texttt{softBlock} list.
\end{itemize}

To enable a scenario object to assign different \texttt{requestPriority} or \texttt{blockPriority} values to different events, we introduced the concept of multi-sync statements (this is a technical solution, to allow different priorities for individual events within the constraints of the BPPy framework). A multi-sync statement allows for multiple individual sync-statements to be used within a single yield point. Each individual sync-statement can have its own \texttt{waitFor}, \texttt{request}, \texttt{softBlock}, and \texttt{hardBlock} lists, as well as unique \texttt{requestPriority} and \texttt{blockPriority} values.

The scenario object will continue past the yield point if the selected event is present in the \texttt{request} or \texttt{waitFor} lists of any of the individual sync-statements in the multi-sync statement at that yield point.

Consider the following example where we are managing a red vertical and diagonal intersection in the game. We need to ensure that the intersection slot is prioritized over the rest of the slots, either for blocking or placing a disc. By using two different sync statements, the intersection slot is given a higher priority, and blocking the slot below is also prioritized accordingly. This guarantees that if the intersection slot is available, it will be selected before any other move, effectively preventing a potential fork. Additionally, the use of soft blocking, rather than the original hard blocking, ensures that the slots under the potential fork are not blocked if there is a higher priority request to place a disc in those positions.

\begin{lstlisting}
@bp.thread
def manage_red_vertical_diagonal_intersection
  (vertical_line, diag_line):
  """
  Monitors intersecting vertical and diagonal lines
  for potential Red fork situations.

  Args:
    vertical_line: Vertical line of positions
                   [(row, col), ...]
    diag_line: Diagonal line of positions 
               [(row, col), ...]
  """

  ...

    
  # Define the synchronization statements
  sync_statements = [
    {
      'waitFor': wait_for,
      'request': request_events,
      'softBlock': block_events,
      'requestPriority': req_priority,
      'blockPriority': block_priority
    }
  ]

  # If the intersection has not been placed, 
  # we request placement with higher priority
  if not intersection_placed:
    sync_statements.append({
      'request': [place_yellow(intersection_slot[0], 
                  intersection_slot[1])],
      'softBlock': below_intersection_slot,
      'requestPriority': intersection_priority,
      'blockPriority': intersection_block_priority
    })

  # Execute multi-sync with the defined sync statements
  event = yield bp.sync(multiSync=sync_statements)

  ...
\end{lstlisting}

\section*{Acknowledgements}
This work was partially funded by the European Union (RobustifAI project, ID 101212818). Views and opinions expressed are however those of the author(s) only and do not necessarily reflect those of the European Union or the European Health and Digital Executive Agency (HADEA). Neither the European Union nor the granting authority can be held responsible for them.

\bibliographystyle{ACM-Reference-Format}
\bibliography{referances.bib}


\begin{thebibliography}{50}


\ifx \showCODEN    \undefined \def \showCODEN     #1{\unskip}     \fi
\ifx \showISBNx    \undefined \def \showISBNx     #1{\unskip}     \fi
\ifx \showISBNxiii \undefined \def \showISBNxiii  #1{\unskip}     \fi
\ifx \showISSN     \undefined \def \showISSN      #1{\unskip}     \fi
\ifx \showLCCN     \undefined \def \showLCCN      #1{\unskip}     \fi
\ifx \shownote     \undefined \def \shownote      #1{#1}          \fi
\ifx \showarticletitle \undefined \def \showarticletitle #1{#1}   \fi
\ifx \showURL      \undefined \def \showURL       {\relax}        \fi
\providecommand\bibfield[2]{#2}
\providecommand\bibinfo[2]{#2}
\providecommand\natexlab[1]{#1}
\providecommand\showeprint[2][]{arXiv:#2}

\bibitem[Allis(1988)]%
        {Al88}
\bibfield{author}{\bibinfo{person}{V. Allis}.} \bibinfo{year}{1988}\natexlab{}.
\newblock \emph{\bibinfo{title}{A Knowledge-Based Approach of Connect-Four ---
  The Game is Solved: White Wins}}.
\newblock \bibinfo{thesistype}{Master's\ thesis}. \bibinfo{school}{Vrije
  Universiteit Amsterdam}.
\newblock


\bibitem[Allis(2025)]%
        {Al25}
\bibfield{author}{\bibinfo{person}{V. Allis}.} \bibinfo{year}{2025}\natexlab{}.
\newblock \bibinfo{title}{{How to Win Connect 4}}.
\newblock
\newblock
\shownote{Technical Report.
  \url{https://www.rd.com/article/how-to-win-connect-4/}}.


\bibitem[Anthropic(2023)]%
        {An23}
\bibfield{author}{\bibinfo{person}{Anthropic}.}
  \bibinfo{year}{2023}\natexlab{}.
\newblock \bibinfo{title}{{Introducing Claude}}.
\newblock
  \bibinfo{howpublished}{\url{https://www.anthropic.com/index/introducing-claude}}.
\newblock


\bibitem[Ashrov and Katz(2023)]%
        {AsKa23}
\bibfield{author}{\bibinfo{person}{A. Ashrov} {and} \bibinfo{person}{G. Katz}.}
  \bibinfo{year}{2023}\natexlab{}.
\newblock \showarticletitle{{Enhancing Deep Learning with Scenario-Based
  Override Rules: a Case Study}}. In \bibinfo{booktitle}{\emph{Proc. 11th Int.
  Conf. on Model-Driven Engineering and Software Development (MODELSWARD)}}.
  \bibinfo{pages}{253--268}.
\newblock


\bibitem[Ashrov et~al\mbox{.}(2025)]%
        {AhStElKa25}
\bibfield{author}{\bibinfo{person}{A. Ashrov}, \bibinfo{person}{A. Sturm},
  \bibinfo{person}{A. Elyasaf}, {and} \bibinfo{person}{G. Katz}.}
  \bibinfo{year}{2025}\natexlab{}.
\newblock \showarticletitle{{A Study on the Comprehensibility of Behavioral
  Programming Variants}}. In \bibinfo{booktitle}{\emph{Proc. 20th Int. Conf. on
  Evaluation of Novel Approaches to Software Engineering (ENASE)}}.
  \bibinfo{pages}{252--267}.
\newblock


\bibitem[Chowdhery et~al\mbox{.}(2022)]%
        {chowdhery2022palm}
\bibfield{author}{\bibinfo{person}{A. Chowdhery}, \bibinfo{person}{S. Narang},
  \bibinfo{person}{J. Devlin}, {et~al\mbox{.}}}
  \bibinfo{year}{2022}\natexlab{}.
\newblock \bibinfo{title}{{PaLM: Scaling Language Modeling with Pathways}}.
\newblock
\newblock
\shownote{Technical Report. \url{https://arxiv.org/abs/2204.02311}}.


\bibitem[Corsi et~al\mbox{.}(2024)]%
        {CoYeAmFaHaKa24}
\bibfield{author}{\bibinfo{person}{D. Corsi}, \bibinfo{person}{R. Yerushalmi},
  \bibinfo{person}{G. Amir}, \bibinfo{person}{A. Farinelli},
  \bibinfo{person}{D. Harel}, {and} \bibinfo{person}{G. Katz}.}
  \bibinfo{year}{2024}\natexlab{}.
\newblock \showarticletitle{{Enforcing Specific Behaviours via Constrained DRL
  and Scenario-Based Programming}}. In \bibinfo{booktitle}{\emph{Proc. 31st
  Int. Conf. on Neural Information Processing (ICONIP)}}.
  \bibinfo{pages}{284--302}.
\newblock


\bibitem[Cursor(2023)]%
        {cursorIDE}
\bibfield{author}{\bibinfo{person}{Cursor}.} \bibinfo{year}{2023}\natexlab{}.
\newblock \bibinfo{title}{{Cursor: The IDE that helps you Code with AI}}.
\newblock \bibinfo{howpublished}{\url{https://www.cursor.com}}.
\newblock


\bibitem[Doe(2025)]%
        {Do25}
\bibfield{author}{\bibinfo{person}{J. Doe}.} \bibinfo{year}{2025}\natexlab{}.
\newblock \bibinfo{title}{Unbeatable AI --- Connect4}.
\newblock
\newblock
\shownote{\url{https://websim.com/@Ch13fB1gT4lk/connect-4-unbeatable-ai}}.


\bibitem[Du et~al\mbox{.}(2024)]%
        {DuLiWaWaLiChFeShPeLo24}
\bibfield{author}{\bibinfo{person}{X. Du}, \bibinfo{person}{M. Liu},
  \bibinfo{person}{K. Wang}, \bibinfo{person}{H. Wang}, \bibinfo{person}{J.
  Liu}, \bibinfo{person}{Y. Chen}, \bibinfo{person}{J. Feng},
  \bibinfo{person}{C. Sha}, \bibinfo{person}{X. Peng}, {and}
  \bibinfo{person}{Y. Lou}.} \bibinfo{year}{2024}\natexlab{}.
\newblock \showarticletitle{{Evaluating Large Language Models in Class-Level
  Code Generation}}. In \bibinfo{booktitle}{\emph{Proc. 46th Int. Conf. on
  Software Engineering (ICSE)}}.
\newblock


\bibitem[Galli(2025)]%
        {Ga25}
\bibfield{author}{\bibinfo{person}{K. Galli}.} \bibinfo{year}{2025}\natexlab{}.
\newblock \bibinfo{title}{{Connect4 Python GitHub Repository}}.
\newblock
\newblock
\shownote{\url{https://github.com/KeithGalli}}.


\bibitem[Greenyer et~al\mbox{.}(2017)]%
        {GeGrGuKoGlMaKa17}
\bibfield{author}{\bibinfo{person}{J. Greenyer}, \bibinfo{person}{D. Gritzner},
  \bibinfo{person}{T. Gutjahr}, \bibinfo{person}{F. K\"{o}nig},
  \bibinfo{person}{N. Glade}, \bibinfo{person}{A. Marron}, {and}
  \bibinfo{person}{G. Katz}.} \bibinfo{year}{2017}\natexlab{}.
\newblock \showarticletitle{{ScenarioTools --- A Tool Suite for the
  Scenario-based Modeling and Analysis of Reactive Systems}}.
\newblock \bibinfo{journal}{\emph{Journal of Science of Computer Programming
  (J. SCP)}}  \bibinfo{volume}{149} (\bibinfo{year}{2017}),
  \bibinfo{pages}{15--27}.
\newblock


\bibitem[Greenyer et~al\mbox{.}(2016a)]%
        {GrGrKaMa16}
\bibfield{author}{\bibinfo{person}{J. Greenyer}, \bibinfo{person}{D. Gritzner},
  \bibinfo{person}{G. Katz}, {and} \bibinfo{person}{A. Marron}.}
  \bibinfo{year}{2016}\natexlab{a}.
\newblock \showarticletitle{{Scenario-Based Modeling and Synthesis for Reactive
  Systems with Dynamic System Structure in ScenarioTools}}. In
  \bibinfo{booktitle}{\emph{Proc. 19th ACM/IEEE Int. Conf. on Model Driven
  Engineering Languages and Systems (MODELS)}}. \bibinfo{pages}{16--23}.
\newblock


\bibitem[Greenyer et~al\mbox{.}(2016b)]%
        {GrGrKaMaGlGuKo16}
\bibfield{author}{\bibinfo{person}{J. Greenyer}, \bibinfo{person}{D. Gritzner},
  \bibinfo{person}{G. Katz}, \bibinfo{person}{A. Marron}, \bibinfo{person}{N.
  Glade}, \bibinfo{person}{T. Gutjahr}, {and} \bibinfo{person}{F. K\"{o}nig}.}
  \bibinfo{year}{2016}\natexlab{b}.
\newblock \showarticletitle{{Distributed Execution of Scenario-Based
  Specifications of Structurally Dynamic Cyber-Physical Systems}}. In
  \bibinfo{booktitle}{\emph{Proc. 3rd Int. Conf. on System-Integrated
  Intelligence: New Challenges for Product and Production Engineering
  (SYSINT)}}. \bibinfo{pages}{552--559}.
\newblock


\bibitem[Haque(2025)]%
        {Ha25}
\bibfield{author}{\bibinfo{person}{M. Haque}.} \bibinfo{year}{2025}\natexlab{}.
\newblock \bibinfo{title}{{LLMs: A Game-Changer for Software Engineers?}}
\newblock
\newblock
\shownote{Technical Report. \url{https://arxiv.org/abs/2411.00932}}.


\bibitem[Harel et~al\mbox{.}(2013a)]%
        {HaKaKa13}
\bibfield{author}{\bibinfo{person}{D. Harel}, \bibinfo{person}{A. Kantor},
  {and} \bibinfo{person}{G. Katz}.} \bibinfo{year}{2013}\natexlab{a}.
\newblock \showarticletitle{{Relaxing Synchronization Constraints in Behavioral
  Programs}}. In \bibinfo{booktitle}{\emph{Proc. 19th Int. Conf. on Logic for
  Programming, Artificial Intelligence and Reasoning (LPAR)}}.
  \bibinfo{pages}{355--372}.
\newblock


\bibitem[Harel et~al\mbox{.}(2013b)]%
        {HaKaKaMaMiWe13}
\bibfield{author}{\bibinfo{person}{D. Harel}, \bibinfo{person}{A. Kantor},
  \bibinfo{person}{G. Katz}, \bibinfo{person}{A. Marron}, \bibinfo{person}{L.
  Mizrahi}, {and} \bibinfo{person}{G. Weiss}.}
  \bibinfo{year}{2013}\natexlab{b}.
\newblock \showarticletitle{{On Composing and Proving the Correctness of
  Reactive Behavior}}. In \bibinfo{booktitle}{\emph{Proc. 13th Int. Conf. on
  Embedded Software (EMSOFT)}}. \bibinfo{pages}{1--10}.
\newblock


\bibitem[Harel et~al\mbox{.}(2015a)]%
        {HaKaKaMaWeWi15}
\bibfield{author}{\bibinfo{person}{D. Harel}, \bibinfo{person}{A. Kantor},
  \bibinfo{person}{G. Katz}, \bibinfo{person}{A. Marron}, \bibinfo{person}{G.
  Weiss}, {and} \bibinfo{person}{G. Wiener}.} \bibinfo{year}{2015}\natexlab{a}.
\newblock \showarticletitle{{Towards Behavioral Programming in Distributed
  Architectures}}.
\newblock \bibinfo{journal}{\emph{Journal of Science of Computer Programming
  (J. SCP)}}  \bibinfo{volume}{98} (\bibinfo{year}{2015}),
  \bibinfo{pages}{233--267}.
\newblock


\bibitem[Harel and Katz(2014)]%
        {HaKa14}
\bibfield{author}{\bibinfo{person}{D. Harel} {and} \bibinfo{person}{G. Katz}.}
  \bibinfo{year}{2014}\natexlab{}.
\newblock \showarticletitle{{Scaling-Up Behavioral Programming: Steps from
  Basic Principles to Application Architectures}}. In
  \bibinfo{booktitle}{\emph{Proc. 4th SPLASH Workshop on Programming based on
  Actors, Agents and Decentralized Control (AGERE!)}}.
  \bibinfo{pages}{95--108}.
\newblock


\bibitem[Harel et~al\mbox{.}(2015b)]%
        {HaKaLaMaWe15}
\bibfield{author}{\bibinfo{person}{D. Harel}, \bibinfo{person}{G. Katz},
  \bibinfo{person}{R. Lampert}, \bibinfo{person}{A. Marron}, {and}
  \bibinfo{person}{G. Weiss}.} \bibinfo{year}{2015}\natexlab{b}.
\newblock \showarticletitle{{On the Succinctness of Idioms for Concurrent
  Programming}}. In \bibinfo{booktitle}{\emph{Proc. 26th Int. Conf. on
  Concurrency Theory (CONCUR)}}. \bibinfo{pages}{85--99}.
\newblock


\bibitem[Harel et~al\mbox{.}(2016a)]%
        {HaKaMaMa16}
\bibfield{author}{\bibinfo{person}{D. Harel}, \bibinfo{person}{G. Katz},
  \bibinfo{person}{R. Marelly}, {and} \bibinfo{person}{A. Marron}.}
  \bibinfo{year}{2016}\natexlab{a}.
\newblock \showarticletitle{{An Initial Wise Development Environment for
  Behavioral Models}}. In \bibinfo{booktitle}{\emph{Proc. 4th Int. Conf. on
  Model-Driven Engineering and Software Development (MODELSWARD)}}.
  \bibinfo{pages}{600--612}.
\newblock


\bibitem[Harel et~al\mbox{.}(2016b)]%
        {HaKaMaMa16b}
\bibfield{author}{\bibinfo{person}{D. Harel}, \bibinfo{person}{G. Katz},
  \bibinfo{person}{R. Marelly}, {and} \bibinfo{person}{A. Marron}.}
  \bibinfo{year}{2016}\natexlab{b}.
\newblock \showarticletitle{{First Steps Towards a Wise Development Environment
  for Behavioral Models}}.
\newblock \bibinfo{journal}{\emph{Int. Journal of Information System Modeling
  and Design (IJISMD)}}  \bibinfo{volume}{7(3)} (\bibinfo{year}{2016}),
  \bibinfo{pages}{1--22}.
\newblock


\bibitem[Harel et~al\mbox{.}(2018)]%
        {HaKaMaMa18}
\bibfield{author}{\bibinfo{person}{D. Harel}, \bibinfo{person}{G. Katz},
  \bibinfo{person}{R. Marelly}, {and} \bibinfo{person}{A. Marron}.}
  \bibinfo{year}{2018}\natexlab{}.
\newblock \showarticletitle{{Wise Computing: Toward Endowing System Development
  with Proactive Wisdom}}.
\newblock \bibinfo{journal}{\emph{IEEE Computer}}  \bibinfo{volume}{51(2)}
  (\bibinfo{year}{2018}), \bibinfo{pages}{14--26}.
\newblock


\bibitem[Harel et~al\mbox{.}(2020)]%
        {HaKaMaSaWe20}
\bibfield{author}{\bibinfo{person}{D. Harel}, \bibinfo{person}{G. Katz},
  \bibinfo{person}{A. Marron}, \bibinfo{person}{A. Sadon}, {and}
  \bibinfo{person}{G. Weiss}.} \bibinfo{year}{2020}\natexlab{}.
\newblock \showarticletitle{{Executing Scenario-Based Specification with
  Dynamic Generation of Rich Events}}.
\newblock \bibinfo{journal}{\emph{Communications in Computer and Information
  Science (CCIS)}}  \bibinfo{volume}{1161} (\bibinfo{year}{2020}),
  \bibinfo{pages}{246--274}.
\newblock


\bibitem[Harel et~al\mbox{.}(2024)]%
        {HaKaMaSz24}
\bibfield{author}{\bibinfo{person}{D. Harel}, \bibinfo{person}{G. Katz},
  \bibinfo{person}{A. Marron}, {and} \bibinfo{person}{S. Szekely}.}
  \bibinfo{year}{2024}\natexlab{}.
\newblock \showarticletitle{{On Augmenting Scenario-Based Modeling with
  Generative AI}}. In \bibinfo{booktitle}{\emph{Proc. 12th Int. Conf. on
  Model-Driven Engineering and Software Development (MODELSWARD)}}.
  \bibinfo{pages}{235--246}.
\newblock


\bibitem[Harel et~al\mbox{.}(2025)]%
        {HaKaMaSz25}
\bibfield{author}{\bibinfo{person}{D. Harel}, \bibinfo{person}{G. Katz},
  \bibinfo{person}{A. Marron}, {and} \bibinfo{person}{S. Szekely}.}
  \bibinfo{year}{2025}\natexlab{}.
\newblock \showarticletitle{{Enhancing Scenario-Based Modeling using Large
  Language Models}}.
\newblock \bibinfo{journal}{\emph{Communications in Computer and Information
  Science (CCIS)}}  \bibinfo{volume}{2547} (\bibinfo{year}{2025}),
  \bibinfo{pages}{43--68}.
\newblock


\bibitem[Harel et~al\mbox{.}(2012a)]%
        {HaKaMaWe12}
\bibfield{author}{\bibinfo{person}{D. Harel}, \bibinfo{person}{G. Katz},
  \bibinfo{person}{A. Marron}, {and} \bibinfo{person}{G. Weiss}.}
  \bibinfo{year}{2012}\natexlab{a}.
\newblock \showarticletitle{{Non-Intrusive Repair of Reactive Programs}}. In
  \bibinfo{booktitle}{\emph{Proc. 17th IEEE Int. Conf. on Engineering of
  Complex Computer Systems (ICECCS)}}. \bibinfo{pages}{3--12}.
\newblock


\bibitem[Harel et~al\mbox{.}(2014)]%
        {HaKaMaWe14}
\bibfield{author}{\bibinfo{person}{D. Harel}, \bibinfo{person}{G. Katz},
  \bibinfo{person}{A. Marron}, {and} \bibinfo{person}{G. Weiss}.}
  \bibinfo{year}{2014}\natexlab{}.
\newblock \showarticletitle{{Non-Intrusive Repair of Safety and Liveness
  Violations in Reactive Programs}}.
\newblock \bibinfo{journal}{\emph{Transactions on Computational Collective
  Intelligence (TCCI)}}  \bibinfo{volume}{16} (\bibinfo{year}{2014}),
  \bibinfo{pages}{1--33}.
\newblock


\bibitem[Harel et~al\mbox{.}(2015c)]%
        {HaKaMaWe15}
\bibfield{author}{\bibinfo{person}{D. Harel}, \bibinfo{person}{G. Katz},
  \bibinfo{person}{A. Marron}, {and} \bibinfo{person}{G. Weiss}.}
  \bibinfo{year}{2015}\natexlab{c}.
\newblock \showarticletitle{{The Effect of Concurrent Programming Idioms on
  Verification}}. In \bibinfo{booktitle}{\emph{Proc. 3rd Int. Conf. on
  Model-Driven Engineering and Software Development (MODELSWARD)}}.
  \bibinfo{pages}{363--369}.
\newblock


\bibitem[Harel et~al\mbox{.}(2011)]%
        {HaLaMaWe11}
\bibfield{author}{\bibinfo{person}{D. Harel}, \bibinfo{person}{R. Lampert},
  \bibinfo{person}{A. Marron}, {and} \bibinfo{person}{G. Weiss}.}
  \bibinfo{year}{2011}\natexlab{}.
\newblock \showarticletitle{{Model-Checking Behavioral Programs}}. In
  \bibinfo{booktitle}{\emph{Proc. 11th Int. Conf. on Embedded Software
  (EMSOFT)}}. \bibinfo{pages}{279--288}.
\newblock


\bibitem[Harel et~al\mbox{.}(2012b)]%
        {HaMaWe12}
\bibfield{author}{\bibinfo{person}{D. Harel}, \bibinfo{person}{A. Marron},
  {and} \bibinfo{person}{G. Weiss}.} \bibinfo{year}{2012}\natexlab{b}.
\newblock \showarticletitle{{Behavioral Programming}}.
\newblock \bibinfo{journal}{\emph{Commun. ACM}} \bibinfo{volume}{55},
  \bibinfo{number}{7} (\bibinfo{year}{2012}), \bibinfo{pages}{90--100}.
\newblock


\bibitem[Katz(2013)]%
        {Ka13}
\bibfield{author}{\bibinfo{person}{G. Katz}.} \bibinfo{year}{2013}\natexlab{}.
\newblock \showarticletitle{{On Module-Based Abstraction and Repair of
  Behavioral Programs}}. In \bibinfo{booktitle}{\emph{Proc. 19th Int. Conf. on
  Logic for Programming, Artificial Intelligence and Reasoning (LPAR)}}.
  \bibinfo{pages}{518--535}.
\newblock


\bibitem[Katz(2020)]%
        {Ka20}
\bibfield{author}{\bibinfo{person}{G. Katz}.} \bibinfo{year}{2020}\natexlab{}.
\newblock \showarticletitle{{Guarded Deep Learning using Scenario-Based
  Modeling}}. In \bibinfo{booktitle}{\emph{Proc. 8th Int. Conf. on Model-Driven
  Engineering and Software Development (MODELSWARD)}}.
  \bibinfo{pages}{126--136}.
\newblock


\bibitem[Katz(2021a)]%
        {Ka21b}
\bibfield{author}{\bibinfo{person}{G. Katz}.} \bibinfo{year}{2021}\natexlab{a}.
\newblock \showarticletitle{{Augmenting Deep Neural Networks with
  Scenario-Based Guard Rules}}.
\newblock \bibinfo{journal}{\emph{Communications in Computer and Information
  Science (CCIS)}}  \bibinfo{volume}{1361} (\bibinfo{year}{2021}),
  \bibinfo{pages}{147--172}.
\newblock


\bibitem[Katz(2021b)]%
        {Ka21}
\bibfield{author}{\bibinfo{person}{G. Katz}.} \bibinfo{year}{2021}\natexlab{b}.
\newblock \showarticletitle{{Towards Repairing Scenario-Based Models with Rich
  Events}}. In \bibinfo{booktitle}{\emph{Proc. 9th Int. Conf. on Model-Driven
  Engineering and Software Development (MODELSWARD)}}.
  \bibinfo{pages}{362--372}.
\newblock


\bibitem[Katz et~al\mbox{.}(2015)]%
        {KaBaHa15}
\bibfield{author}{\bibinfo{person}{G. Katz}, \bibinfo{person}{C. Barrett},
  {and} \bibinfo{person}{D. Harel}.} \bibinfo{year}{2015}\natexlab{}.
\newblock \showarticletitle{{Theory-Aided Model Checking of Concurrent
  Transition Systems}}. In \bibinfo{booktitle}{\emph{Proc. 15th Int. Conf. on
  Formal Methods in Computer-Aided Design (FMCAD)}}. \bibinfo{pages}{81--88}.
\newblock


\bibitem[Katz and Elyasaf(2021)]%
        {KaEl21}
\bibfield{author}{\bibinfo{person}{G. Katz} {and} \bibinfo{person}{A.
  Elyasaf}.} \bibinfo{year}{2021}\natexlab{}.
\newblock \showarticletitle{{Towards Combining Deep Learning, Verification, and
  Scenario-Based Programming}}. In \bibinfo{booktitle}{\emph{Proc. 1st Workshop
  on Verification of Autonomous and Robotic Systems (VARS)}}.
  \bibinfo{pages}{1--3}.
\newblock


\bibitem[Katz et~al\mbox{.}(2019)]%
        {KaMaSaWe19}
\bibfield{author}{\bibinfo{person}{G. Katz}, \bibinfo{person}{A. Marron},
  \bibinfo{person}{A. Sadon}, {and} \bibinfo{person}{G. Weiss}.}
  \bibinfo{year}{2019}\natexlab{}.
\newblock \showarticletitle{{On-the-Fly Construction of Composite Events in
  Scenario-Based Modeling Using Constraint Solvers}}. In
  \bibinfo{booktitle}{\emph{Proc. 7th Int. Conf. on Model-Driven Engineering
  and Software Development (MODELSWARD)}}. \bibinfo{pages}{143--156}.
\newblock


\bibitem[Li et~al\mbox{.}(2024)]%
        {LiPaPo24}
\bibfield{author}{\bibinfo{person}{Y. Li}, \bibinfo{person}{J. Parsert}, {and}
  \bibinfo{person}{E. Polgreen}.} \bibinfo{year}{2024}\natexlab{}.
\newblock \showarticletitle{{Guiding Enumerative Program Synthesis with Large
  Language Models}}. \bibinfo{publisher}{Proc. 36th Int. Conf. on Computer
  Aided Verification (CAV)}.
\newblock


\bibitem[Marron et~al\mbox{.}(2018)]%
        {MaHaHaMuTe18}
\bibfield{author}{\bibinfo{person}{A. Marron}, \bibinfo{person}{Y. Hacohen},
  \bibinfo{person}{D. Harel}, \bibinfo{person}{A. M{\"u}lder}, {and}
  \bibinfo{person}{A. Terfloth}.} \bibinfo{year}{2018}\natexlab{}.
\newblock \showarticletitle{{Embedding Scenario-based Modeling in
  Statecharts}}. In \bibinfo{booktitle}{\emph{Proc. 5th Int. Workshop on
  Model-driven Robot Software Engineering (MORSE)}}. \bibinfo{pages}{443--452}.
\newblock


\bibitem[OpenAI(2023)]%
        {openai2023gpt4}
\bibfield{author}{\bibinfo{person}{OpenAI}.} \bibinfo{year}{2023}\natexlab{}.
\newblock \bibinfo{title}{{GPT-4}}.
\newblock
\newblock
\shownote{Technical Report. \url{https://openai.com/research/gpt-4}}.


\bibitem[Steinberg et~al\mbox{.}(2017)]%
        {StGrGrHaKaMa17}
\bibfield{author}{\bibinfo{person}{S. Steinberg}, \bibinfo{person}{J.
  Greenyer}, \bibinfo{person}{D. Gritzner}, \bibinfo{person}{D. Harel},
  \bibinfo{person}{G. Katz}, {and} \bibinfo{person}{A. Marron}.}
  \bibinfo{year}{2017}\natexlab{}.
\newblock \showarticletitle{{Distributing Scenario-Based Models: A
  Replicate-and-Project Approach}}. In \bibinfo{booktitle}{\emph{Proc. 5th Int.
  Conf. on Model-Driven Engineering and Software Development (MODELSWARD)}}.
  \bibinfo{pages}{182--195}.
\newblock


\bibitem[Steinberg et~al\mbox{.}(2018)]%
        {StGrGrHaKaMa18}
\bibfield{author}{\bibinfo{person}{S. Steinberg}, \bibinfo{person}{J.
  Greenyer}, \bibinfo{person}{D. Gritzner}, \bibinfo{person}{D. Harel},
  \bibinfo{person}{G. Katz}, {and} \bibinfo{person}{A. Marron}.}
  \bibinfo{year}{2018}\natexlab{}.
\newblock \showarticletitle{{Efficient Distributed Execution of Multi-Component
  Scenario-Based Models}}.
\newblock \bibinfo{journal}{\emph{Communications in Computer and Information
  Science (CCIS)}}  \bibinfo{volume}{880} (\bibinfo{year}{2018}),
  \bibinfo{pages}{449--483}.
\newblock


\bibitem[Sun et~al\mbox{.}(2024)]%
        {SuShPaBa24}
\bibfield{author}{\bibinfo{person}{C. Sun}, \bibinfo{person}{Y. Sheng},
  \bibinfo{person}{O. Padon}, {and} \bibinfo{person}{C. Barrett}.}
  \bibinfo{year}{2024}\natexlab{}.
\newblock \showarticletitle{{Clover: Closed-Loop Verifiable Code Generation}}.
  In \bibinfo{booktitle}{\emph{Proc. 1st Int. Symposium on AI Verification
  (SAIV)}}. \bibinfo{pages}{134--155}.
\newblock


\bibitem[Tleemann(2025)]%
        {Tl25}
\bibfield{author}{\bibinfo{person}{T. Tleemann}.}
  \bibinfo{year}{2025}\natexlab{}.
\newblock \bibinfo{title}{{Pro Player --- Connect4}}.
\newblock
\newblock
\shownote{\url{https://tleemann.de/four.html}}.


\bibitem[Yaacov(2023)]%
        {Ya23}
\bibfield{author}{\bibinfo{person}{T. Yaacov}.}
  \bibinfo{year}{2023}\natexlab{}.
\newblock \showarticletitle{{BPpy: Behavioral Programming in Python}}.
\newblock \bibinfo{journal}{\emph{SoftwareX}}  \bibinfo{volume}{24}
  (\bibinfo{year}{2023}).
\newblock


\bibitem[Yaacov et~al\mbox{.}(2024)]%
        {YaElWe24}
\bibfield{author}{\bibinfo{person}{T. Yaacov}, \bibinfo{person}{A. Elyasaf},
  {and} \bibinfo{person}{G. Weiss}.} \bibinfo{year}{2024}\natexlab{}.
\newblock \showarticletitle{{Boosting LLM-Based Software Generation by Aligning
  Code with Requirements}}. \bibinfo{publisher}{Proc. IEEE 32nd Int.
  Requirements Engineering Conference Workshops (REW)}.
\newblock


\bibitem[Yaacov et~al\mbox{.}(2025)]%
        {YaWeAsKaZi25}
\bibfield{author}{\bibinfo{person}{T. Yaacov}, \bibinfo{person}{G. Weiss},
  \bibinfo{person}{A. Ashrov}, \bibinfo{person}{G. Katz}, {and}
  \bibinfo{person}{H. Zisser}.} \bibinfo{year}{2025}\natexlab{}.
\newblock \showarticletitle{{Exploring and Evaluating Interplays of BPpy with
  Deep Reinforcement Learning and Formal Methods}}. In
  \bibinfo{booktitle}{\emph{Proc. 20th Int. Conf. on Evaluation of Novel
  Approaches to Software Engineering (ENASE)}}. \bibinfo{pages}{27--40}.
\newblock


\bibitem[Yerushalmi et~al\mbox{.}(2022)]%
        {YeAmElHaKaMa22}
\bibfield{author}{\bibinfo{person}{R. Yerushalmi}, \bibinfo{person}{G. Amir},
  \bibinfo{person}{A. Elyasaf}, \bibinfo{person}{D. Harel}, \bibinfo{person}{G.
  Katz}, {and} \bibinfo{person}{A. Marron}.} \bibinfo{year}{2022}\natexlab{}.
\newblock \showarticletitle{{Scenario-Assisted Deep Reinforcement Learning}}.
  In \bibinfo{booktitle}{\emph{Proc. 10th Int. Conf. on Model-Driven
  Engineering and Software Development (MODELSWARD)}}.
  \bibinfo{pages}{310--319}.
\newblock


\bibitem[Yerushalmi et~al\mbox{.}(2023)]%
        {YeAmElHaKaMa23}
\bibfield{author}{\bibinfo{person}{R. Yerushalmi}, \bibinfo{person}{G. Amir},
  \bibinfo{person}{A. Elyasaf}, \bibinfo{person}{D. Harel}, \bibinfo{person}{G.
  Katz}, {and} \bibinfo{person}{A. Marron}.} \bibinfo{year}{2023}\natexlab{}.
\newblock \showarticletitle{{Enhancing Deep Reinforcement Learning with
  Scenario-Based Modeling}}.
\newblock \bibinfo{journal}{\emph{Springer Nature Computer Science (SNCS)}}
  \bibinfo{volume}{4} (\bibinfo{year}{2023}), \bibinfo{pages}{1--13}.
\newblock


\end{thebibliography}

\end{document}